\documentclass[onecollarge]{svjour2}
\journalname{Few Body Systems} 
\usepackage[dvips]{graphicx}
\usepackage{tabls}
\def\ga{\alpha}
\def\gb{\beta}

\def\ge{\epsilon}
\def\gg{\gamma}
\def\gd{\delta}

\def\gm{\mu}
\def\gn{\nu}
\def\gO{\Omega}
\def\gp{\pi}
\def\gP{\Pi}
\def\gr{\rho}

\def\gS{\Sigma}

\def\delmu{\partial_\gm}
\def\delmuu{\partial^\gm}
\def\delnu{\partial_\gn}

\def\delrlmu{\stackrel {\leftrightarrow} {\partial_\gm}}

\def\delrl1{\stackrel {\leftrightarrow} {\partial_1}}

\def\part{\partial}

\def\hlf{\frac{1}{2}}
\def\A0{A^{+}_0}

\def\psid{\psi^\dagger}
\def\Psid{\Psi^\dagger}

\def\psib{\overline{\psi}}

\def\Phib{\overline{\Phi}}

\def\Psib{\overline{\Psi}}

\def\psid{\psi^{\dag}}

\def\ulix{\underline{x}}
\def\uliy{\underline{y}}

\newcommand{\nc}{\newcommand}
\nc{\intl}{\int\limits_{-L}^{+L}\!\frac{{\rm d}x^-}{2}}
\nc{\intly}{\int\limits_{-L}^{+L}\!{{{\rm d}y^-}\over\!2}}
\nc{\intlz}{\int\limits_{-L}^{+L}\!{{{\rm d}z^-}\over\!2}}
\nc{\intlu}{\int\limits_{-L}^{+L}\!{{{\rm d}u^-}\over\!2}}
\nc{\intlv}{\int\limits_{-L}^{+L}\!{{{\rm d}v^-}\over\!2}}
\nc{\intv}{\int\limits_{-V}^{}\!{\rm d}^3\ulix}
\nc{\intvy}{\int\limits_{-V}^{}\!{\rm d}^3\uliy}
\nc{\zmint}{\int\limits_{-L}^{+L}\!{{{\rm d}x^-}\over{\!2L}}}
\nc{\zminty}{\int\limits_{-L}^{+L}\!{{{\rm d}y^-}\over{\!2L}}}
\nc{\intp}{\int\limits_{0}^{+\infty}\!{{{\rm d}p^+}\over\!4\gp}}
\nc{\inp}{\int\limits_{0}^{\infty}\!{{{\rm d}p^+}\over{\!2\sqrt{2\gp}}}}
\nc{\inq}{\int\limits_{0}^{\infty}\!{{{\rm d}q^+}\over{\!2\sqrt{2\gp}}}}
\nc{\inpp}{\int\limits_{0}^{\infty}\!{{{\rm d}p^+}\over{\!2p^+\sqrt{2\gp}}}}
\nc{\ink}{\int\limits_{0}^{\infty}\!{\rm d}k^+}
\nc{\inqq}{\int\limits_{0}^{\infty}\!{{{\rm d}q^+}\over{\!2q^+\sqrt{2\gp}}}}
\nc{\insl}{\int\limits_{-L}^{+L}\!{\rm d}x}
\nc{\intex}{\int\limits_{-\infty}^{+\infty}\!{\rm d}x^1}
\nc{\intey}{\int\limits_{-\infty}^{+\infty}\!{\rm d}y^1}
\nc{\intep}{\int\limits_{-\infty}^{+\infty}\!{\rm d}p^1}
\nc{\inteq}{\int\limits_{-\infty}^{+\infty}\!{\rm d}q^1}
\nc{\intek}{\int\limits_{-\infty}^{+\infty}\!{\rm d}k^1}
\nc{\intel}{\int\limits_{-\infty}^{+\infty}\!{\rm d}l^1}
\nc{\inty}{\int\limits_{-\infty}^{+\infty}\!{\rm d}y^-}
\nc{\intz}{\int\limits_{-\infty}^{+\infty}\!{\rm d}z^-}
\def\beq{\begin{equation}}
\def\eeq{\end{equation}}
\def\bea{\begin{eqnarray}}
\def\eea{\end{eqnarray}}
\nc{\intx}{\int\limits_{-\infty}^{+\infty}\!\frac{{\rm d}x^-}{2}}
\nc{\intgix}{\int\limits_{-\infty}^{+\infty}\!{{{\rm d}x^-}\over\!2}}
\nc{\intgiy}{\int\limits_{-\infty}^{\infty}\!{{{\rm d}y^-}\over\!2}}
\begin{document}
\title{Exactly solvable models 
and spontaneous symmetry breaking
\footnote{Presented by the author at LIGHTCONE 2011, 23--27 May, 2011, Dallas}}
%\medskip
\author{L$\!\!$'ubom\'{\i}r Martinovi\u{c}}
\institute{\at 
Institute of Physics, Slovak Academy of Sciences, 
D\'ubravsk\'a cesta 9, 845 11 Bratislava, Slovakia  \\ 
Tel.:+421-2-5941 055,~ 
Fax: +421-2-5477 6085\\
\email{fyziluma@savba.sk}\\  
\emph{Present address:} BLTP JINR, 141 980 Dubna, Russia 
}
\date{}
\maketitle
\begin{abstract}
We study a few two-dimensional models with massless and massive  
fermions in the hamiltonian framework and in both conventional and light-front  
forms of field theory. The new ingredient is a modification of the canonical 
procedure by taking into account solutions of the operator field equations. 
After summarizing the main results for the derivative-coupling and the 
Thirring models, we briefly compare conventional and light-front versions of 
the Federbush model including the massive current bosonization and a Bogoliubov 
transformation to diagonalize the Hamiltonian. Then we sketch   
an extension of our hamiltonian approach to the two-dimensional  
Nambu -- Jona-Lasinio model and the Thirring-Wess model. Finally, we discuss   
the Schwinger model in a covariant gauge. In particular, we point out  
that the solution due to Lowenstein and Swieca implies the physical 
vacuum in terms of a coherent state of massive scalar field and suggest  
a new formulation of the model's vacuum degeneracy.  
\keywords{solvable relativistic models \and operator solutions \and 
vacuum structure \and Bogoliubov transformation \and spontaneous symmetry 
breaking} 
\end{abstract}
%+++by a Bogoliubov transformation and discuss 
%%%schemes. We suggest it for a detailed nonperturbative comparison between the 
%%%two forms of the relativistic dynamics.  
%\tableofcontents
%***\tc{green}{in collaboration with P. Grang\'e}
\section{Introduction}
\label{intro}

Exactly solvable models may seem to be almost a closed chapter in the 
development of quantum field theory. Our original aim for looking at this 
class of simple relativistic theories was to learn about their properties 
in the light-front (LF) formulation and compare that picture with the 
standard one (by the latter we mean conventional operator formalism in terms   
of usual space-like (SL) field variables). Surprisingly enough, a closer    
look at these models revealed certain inconsistencies and  
contradictions in the known SL solutions. For example, the Fock vacuum   
was taken as the true ground state for calculations of the correlation  
functions in the Thirring model \cite{Klaib} although a derivation of the  
model's Hamiltonian shows that it is non-diagonal when expressed in  
terms of corresponding creation and annihilation operators. Hence the Fock  
vacuum cannot be its (lowest-energy) eigenstate. Another example is the  
(massless) Schwinger model, often invoked as a prototype for more  
complicated gauge theories. It turns out that the widely-accepted  
covariant-gauge solution \cite{LSw} involves some unphysical degrees of  
freedom as a consequence of residual gauge freedom. A rigorous analysis  
\cite{MStr} in the axiomatic spirit showing some spurious features of the 
solution \cite{LSw} remained almost unnoticed and a less formal, e.g. 
a hamiltonian study correcting the physical picture seems to be lacking in 
literature.    

Defining property of the soluble models is that one can write down  
operator solutions of the field equations. Hence, one should be able to   
extract their physical content completely. A novel feature that  
has been overlooked so far is a necessity to formulate these models in terms  
of true field degrees of freedom -- the free fields. As we show below, the  
operator solutions are always composed from free fields. One 
should take this information into account by re-expressing the Lagrangiam and  
consequently the Hamiltonian in terms of these variables. One can  
summarize the above statement by a slogan: "work with the right Hamiltonian!" 
Recently, we applied this strategy to the massive derivative-coupling  
model (DCM), the Thirring and the Federbush 
models. The brief summary of the achieved results is given in the following 
section. Then we extend our approach to the case of the chiral Gross-Neveu  
model, the Thirring-Wess and the Schwinger model. Not all the details are 
worked out at the present stage -- we formulate the main ideas and 
indicate the strategy to be followed. The full treatment of the models will 
be given separately \cite{lmpg}. We conclude the present paper with a brief   
description of the symmetry-breaking pattern fully based on the light-front 
dynamics.  

\section{Hamiltonian study of three models - DCM, Thirring and Federbush}
\label{sec:1}
{\bf The model with derivative coupling} \cite{Schro} 
turns out to be almost a trivial one. 
Its Lagrangian    
\beq
{\cal L}=\frac{i}{2}\Psib\gg^\mu\delrlmu \Psi - m\Psib\Psi + 
\hlf \delmu \phi 
\delmuu \phi - \hlf \mu^2 \phi^2 - g\delmu \phi J^\mu,~~
J^\mu=\Psib \gg^\mu \Psi.
\label{DL}
\eeq
%**For $\mu=0$ known as the Schroer's model, for axial vector current 
%**interaction as Rothe-Stamatescu model ($m = 0, \mu \neq 0$). 
%**i\gg^\mu\delmu \Psi = m\Psi + g\delmu \phi\gg^\mu \Psi, \nonumber \\
%**\delmu \delmuu \phi + \mu^2 \phi = g\delmu J^\mu = 0.
%**\tc{green}{Convention:}~capital Greek letters - interacting Heisenberg 
%**fields, small - free fields
leads to the Dirac equation which is explicitly solved in terms of a 
free scalar and fermion field:  
\beq
\Psi(x) = :e^{ig\phi(x)}:\psi(x),~~i\gg^\mu\delmu \psi(x) =  m\psi(x).
\label{dcmsol}
\eeq
The massive scalar field $\phi(x)$ obeys the free Klein-Gordon equation as 
a consequence of the current conservation.  
The conventional canonical treatment yields a surprising result: the 
obtained LF Hamiltonian is a free one while the SL Hamiltonian contains an 
interacting piece, which is non-diagonal in terms of Fock operators and 
hence its true ground state (which can be obtained by a Bogoliubov 
transformation) differs from the Fock vacuum. So the physical pictures in two  
quantization schemes contradict each other. The explanation is simple.      
One observes that the solution (\ref{dcmsol}) means that there is no 
independent interacting field - it is composed from the free 
fields. We have to insert the solution to the Lagrangian first (analogously 
to inserting a constraint into a Lagrangian), then calculate conjugate momenta 
and derive the Hamiltonian. In this way, a free Lagrangian and Hamiltonian are 
found also in the SL case. The new procedure does not alter the LF result. 
The correlation functions in the two schemes 
coincide as well. They are built from free scalar and fermion two-point 
functions.       

{\bf The Thirring model} \cite{Thir} with its Lagrangian describing a   
self-interacting massless Fermi field  
%%first solved by Thirring by Bethe Ansatz, K. Johson found and 
%%solved a coupled set of equations for the Green's functions (based on 
%%conservation of vector and axial vector currents)
%%Thirring model may seem obsolete and uninteresting today 

%%{\tc{green}{Question:}} why is it necessary to make Ansaetze or 
%%apply approximative methods when we know the exact solution and can 
%%calculate all properties, including vacuum state, directly? 
 
%**systematic Hamiltonian study based on the model's solvability not 
%**given so far

\beq
{\cal L}=\frac{i}{2}\Psib\gg^\mu\delrlmu \Psi - \hlf gJ_\mu J^\mu,~~
J^\mu=\Psib \gg^\mu \Psi
\label{TL}
\eeq
%**Field equations and current conservation
%**&&i\gg^\mu\partial_\mu \Psi(x) = gJ^\mu(x)\gg_\mu\Psi(x),\nonumber \\
%**&&\partial_\mu J^\mu(x) = 0.
is a more complicated theory. The simplest solution is similar to  
Eq.(\ref{dcmsol}) but the elementary scalar field $\phi$ is replaced by the 
composite field $j(x)$ defined via   
$J_\mu = j_\mu = \frac{1}{\sqrt{\pi}}\partial_\mu j$.
The corresponding Hamiltonian $H$ is non-diagonal in composite boson  
operators $c, c^\dagger$ built from fermion bilinears according to 
\bea 
&&\!\!\!\!\!\!\!j^\mu(x) = -\frac{i}{\sqrt{2}\pi}\int\frac{dk^1}{\sqrt{2k^0}}k^\mu
\big\{c(k^1)e^{-i\hat{k}.x} - c^\dagger(k^1)e^{i\hat{k}.x}\big\}, \\ 
&&\!\!\!\!\!\!\!c(k^1) = \frac{i}{\sqrt{k^0}}\int dp^1\big\{\theta\big(p^1k^1\big)
\big[b^\dagger(p^1)b(p^1+k^1) - (b \rightarrow d)\big] + 
%+++d^\dagger(p^1)d(p^1+k^1)\big] + 
\ge(p^1)\theta\big(p^1(p^1-k^1)\big)d(k^1-p^1)b(p^1)\big\}.
\label{bcur}
\eea  
A diagonalization by a Bogoliubov transformation $UHU^{-1}$ implemented by 
a unitary operator $U[\gg(g)]$, where $\gg$ is a suitably chosen function, 
generates the true ground state as  
\beq
\vert \gO \rangle = 
%%e^{-\gb(\gg_D)}
N \exp\big\{-\kappa\intep c^\dagger(p^1)
c^\dagger(-p^1)\big\}
\vert 0 \rangle.
\label{vacsim}
\eeq
Here $N$ is a normalization factor and $\kappa$ is a $g$-dependent function. 
$\vert \gO\rangle$ corresponds to a coherent state of pairs of composite bosons 
with zero values of the total momentum, charge and axial charge. Thus  
no chiral symmetry breaking occurs in the model at least for  
$g \leq \gp$ where the diagonalization is valid.  
%+++P^1 \vert \gO \rangle = 0,~~P^1=\intep p^1 \big[b^\dagger(p^1)b(p^1) + 
%+++d^\dagger(p^1)d(p^1)\big].  
%+++\label{vacmom}
%+++The vacuum $\vert \gO \rangle$ is invariant under $U(1)$ and $U_A(1)$ 
%+++transformations (i.e. carries vanishing charge and axial charge):  
%+++&&\!\!\!\!U(\ga)\vert \gO\rangle = \vert \gO \rangle,~~U(\ga) = e^{i\ga Q}, 
%+++~~Q = \inteq \big[b^\dagger(q)b(q) - d^\dagger(q)d(q)\big],  
%+++\nonumber \\  
%+++&&\!\!\!\!V(\gb)\vert \gO\rangle = \vert \gO \rangle,~~V(\gb) = e^{i\gb 
%+++Q_5},~~Q_5 = \inteq \ge(q)\big[b^\dagger(q)b(q) - d^\dagger(q)d(q)\big].
%+++\label{QQ}
%+++\nonumber 
%+++The vacuum state $\vert \gO \rangle$ corresponds to the symmetric phase,  
%+++no chiral symmetry breaking 
%**-- in contradiction with the results of Faber and Ivanov 
%%who used a Nambu--Jona-Lasinio type of ansatz for the vacuum state claiming 
%%that it is energetically favourable for the theory to exist in the broken 
%%phase. \tc{green}{Problems:} an ansatz for an exactly soluble model? 
%**true vacuum should be 
%**an eigenstate of the full Hamiltonian! $\vert \gO \rangle$ is such a state 
%%where is a phase transition in Faber and Ivanov's approach ??   
%%%%\newpage

{\bf The Federbush model} \cite{Feder}  
is the only known {\it massive} solvable model. The solvability comes from a  
specific current-current coupling between two species of 
massive fermions described by the Lagrangian 
%+++in a manner paralel to above Thirring model case requires a generalization 
%+++of the Klaiber's bosonization (\ref{bcur}) to  massive fermions. Then one 
%+++can search for the true physical ground state by means of a massive 
%+++version of the Bogoliubov transformation.  
\beq  
{\cal L}=\frac{i}{2}\Psib\gg^\mu\delrlmu \Psi - m\Psib\Psi +  
 \frac{i}{2}\Phib\gg^\mu\delrlmu \Phi - \mu\Phib\Phi - g\ge_{\mu\nu}J^\mu
H^\nu. 
%%\ge^{\mu\nu}=-\ge^{\nu\mu},~~~\ge^{01}=1 \Rightarrow \ge^{+-} = 2. 
\label{FfL}
\eeq  
Here the currents are $J^\mu =\Psib\gg^\mu\Psi,~H^\mu=\Phib\gg^\mu\Phi$. 
The coupled field equations
\beq 
i\gg^\mu\delmu\Psi(x) = m\Psi(x) + g\ge_{\mu\nu}\gg^\mu H^\nu(x)\Psi(x),~~~~ 
i\gg^\mu\delmu\Phi(x) = \mu\Phi(x) - g\ge_{\mu\nu}\gg^\mu J^\nu(x)\Phi(x)  
\eeq 
have the solution of the form (\ref{dcmsol}) 
with two "integrated currents" $j(x)$ and $h(x)$.  
%**Unlike the closely related massive Thirring model, Federbush 
%**model is exactly solvable.
%+++The other properties of the model :
%+++\item{Federbush model: exactly solvable with massive fermions (OK for LF)}  
One again finds that the structure of the SL and LF Hamiltonians coincides 
only when the operator solution is implemented in the Lagrangians. However, 
the SL Hamiltonian is not diagonal and a Bogoliubov   
transformation is needed to find the physical ground state. This requires  
a generalization of Klaiber's massless bosonization yielding a    
complicated substitute for $c(k^1)$ of Eq.(\ref{bcur}). In a sharp    
contrast, the LF massive bosonization is as simple as the SL massless one.    
The model is very suitable for a non-perturbative comparison of the two forms 
of the relativistic dynamics. This is because 2-D massless fields 
cannot be treated directly in the LF formalism (only as the massless limits of 
massive theories - this is obvious already from the LF massive 
two-point functions).  
%+++ -- correlation functions from the operator 
%+++solutions using the corresponding vacuum states should coincide}  
%%%
%+++&&i\gg^\mu\delmu\Psi(x) = m\Psi(x) + g\ge_{\mu\nu}\gg^\mu H^\nu(x)\Psi(x) 
%+++\nonumber \\ 
%+++&&i\gg^\mu\delmu\Phi(x) =\mu\Phi(x)- g\ge_{\mu\nu}\gg^\mu J^\nu(x)\Phi(x).  
%+++The "integrated currents" $j(x)$ and $h(x)$ 
%+++J^\mu(x) = \frac{\ge^{\mu\nu}}{\sqrt{\gp}}\delnu j(x),~~ 
%+++H^\mu(x) = \frac{\ge^{\mu\nu}}{\sqrt{\gp}}\delnu h(x), 
%+++enter into the solutions in a "non-diagonal" way:
%+++\Psi(x) = e^{-i\frac{g}{\sqrt{\gp}}h(x)}\psi(x),~~
%+++\Phi(x) = e^{i\frac{g}{\sqrt{\gp}}j(x)}\phi(x).  
Exponentials of the massive composite fields are more singular than the 
massless once. They have to be defined using the "triple-dot ordering" 
\cite{Wigh,SchroT} which  
%+++(Wightman, Schroer)  
generalizes the normal ordering (subtractions of the VEVs  
order by order). We avoid this by bosonization of the massive current. 
The price we pay is complicated commutators at unequal times that are 
needed for computation of correlation functions. In any case, the correlators 
in the SL and LF version of the theory should coincide in form. A remarkable 
albeit for the moment only a conjectured scenario is that this will indeed 
happen with complicated operator structures plus non-trivial vacuum structure 
in the SL case and with much simpler operator part plus the Fock vacuum in the 
LF case.    
\section{Chiral Gross-Neveu model}
\label{sec:2}
The Lagrangian and field equations of the chiral Gross-Neveu model \cite{GN} 
are 
\beq  
{\cal L}=\frac{i}{2}\Psib\gg^\mu\delrlmu \Psi - \frac{g}{2}\big[\big(
\Psib\Psi\big)^2 + \big(\Psib i\gg^5 \Psi\big)^2\big],~~ 
%+++\label{njll}
i\gg^\mu\partial_\mu\Psi = g\big[\big(\Psib\Psi\big)\Psi-\big(\Psib\gg^5\Psi
\big)\gg^5\Psi\big].
\label{njlem}
\eeq
This theory is a 2-D version of the Nambu--Jona-Lasinio model. When rewritten 
in the component form, one realizes that the above equations up to 
the sign coincide with those in the Thirring model:    
\bea 
i\Big(\partial_0 - \partial_1\Big)\Psi_1 = 2g\Psid_2\Psi_1\Psi_2 = 
-2g\Psid_2\Psi_2\Psi_1 = -g(j^0 + j^1), 
\nonumber \\ 
i\Big(\partial_0 + \partial_1\Big)\Psi_2 = 2g\Psid_1\Psi_2\Psi_1 = 
-2g\Psid_1\Psi_1\Psi_2 = -g(j^0 - j^1). 
\label{njleq}
\eea  
Inserting the solution known from the Thirring model into the Lagrangian, 
we find the Hamiltonian 
\beq
H = \intex \Big\{-i\psi^\dagger\ga^1\partial_1 \psi +  
\frac{1}{2}g\big(j^0j^0 - j^1j^1\big) 
 + \frac{g}{2}\Big[\Big(\psib\psi\Big)^2-\Big(\psib\gg^5\psi\Big)^2\Big].  
\eeq
We can bosonize the scalar densities 
$\Sigma(x) = \psib\psi = \psid_1\psi_2 + \psid_2\psi_1$, $\Sigma_5(x) = 
i\psib\gg^5\psi = i\big(\psid_1\psi_2 
- \psid_2\psi_1\big)$: 
%+++ analogously to the current (\ref{bcur}). 
%+++The current-current term can be bosonized as for the Thir model (opposite 
%+++sign, however), the last term bosonized analogously:  
%+++++++++Writing 
\beq   
\Sigma(x) = \int\limits^{+\infty}_{-\infty}\frac{dk^1}{2\gp}
\Big[A(k^1,t)e^{ik^1x^1} + 
A^\dagger(k^1,t)e^{-ik^1x^1}\Big],
~\Sigma_5(x) = \int\limits^{+\infty}_{-\infty}\frac{dk^1}{2\gp}
\Big[B(k^1,t)e^{ik^1x^1} + H.c.\Big] \nonumber 
%++++B^\dagger(k^1,t)e^{-ik^1x^1}\Big], \nonumber 
\label{Fsig}
\eeq
With the Fock expansions for the Fermi field and after the Fourier 
transform, we obtain  
\bea 
&&A(k^1,t) =   
\intep\Big\{\frac{1}{2}\Big[b^\dagger(-p^1)b(k^1-p^1)+d^\dagger
(-p^1)d(k^1-p^1)\Big]\theta\big(p^1(k^1-p^1)\big)e^{i2p^1t} \nonumber \\ 
&&~~~~~~~~~~~~~~~- \theta(p^1k^1)\ge(p^1)\big[d(-p^1)b(p^1+k^1)+
b(-p^1)d(p^1+k^1)\big]
e^{-i2p^1t}\Big\}e^{-ik^1t},    
\label{Ak}
\eea
and similarly for $B(k^1,t)$. One then has to diagonalize the Hamiltonian 
$H=H_0 + H_{1}+H_{2}$ where 
%+++H_0 = \intep \vert p^1 \vert \Big[b^\dagger(p^1)b(p^1) + 
%+++d^\dagger(p^1)d(p^1)\Big].
%+++\label{freh}
\bea 
&&H_1 = - \frac{g}{\gp}\int\limits_{-\infty}^{+\infty} dk^1 \vert k^1 
\vert \Big[
c^\dagger(k^1)c^\dagger(-k^1) + c(k^1)c(-k^1)\Big], \nonumber \\. 
&&H_2 = - \frac{g}{4\gp}\int\limits_{-\infty}^{+\infty} dk^1 \vert k^1 
\vert \Big\{\Big[2A^\dagger(k^1)A(k^1) + 
A^\dagger(k^1)A^\dagger(-k^1) + A(k^1)A(-k^1)\Big] +\Big[A\rightarrow B 
\Big]\Big\}. 
\eea 
Obviously $\vert 0 \rangle$ is not an eigenstate of $H$. The true vacuum has  
to be found by a Bogoliubov transformation. It will be different than 
the Thirring-model vacuum and probably non-invariant under $Q_5$. This  
remains to be verified. We also leave for future work generalization of the 
model to $N_f$ flavours.  

\section{Thirring-Wess model} 
\label{sec:3}
This model \cite{TW} is simpler than the Schwinger model because the nonzero 
bare mass  
of the vector field removes gauge invariance with all its subtleties. The     
corresponding Lagrangian 
\beq
{\cal L}=\frac{i}{2}\Psib\gg^\mu\delrlmu \Psi-\frac{1}{4}G_{\mu\nu}G^{\mu\nu} 
+\mu_0^2B_\mu B^\mu -eJ_\mu B^\mu,~~~G_{\mu\nu}=\delmu B_\nu-
\partial_\nu B_\mu.  
\label{TW}
\eeq
%+++Analyzed by L. Brown, point-splitting with the "gauge-field" exponential
%+++not correct and necessary; 
leads to the field equations -- the Dirac and Proca equations:  
\beq
i\gg^\mu\partial_\mu\Psi(x)= e\gg^\mu B_\mu(x)\Psi(x),~~ 
\delmu G^{\mu\nu}(x) + \mu_0^2 B^\mu(x) = eJ^\mu(x).  
\label{dproc}
\eeq
The rhs of the second equation reduces to $(\partial_\nu\partial^\nu + 
\mu_0^2)B^\nu$  
since $\partial_\nu B^\nu = 0$ due to the current conservation.  
The latter condition also permits us to write down the solution of the 
corresponding Dirac equation as   
\beq
\Psi(x)=\exp\Big\{-\frac{ie}{2}\gg^5\intey \ge(x^1-y^1)B^0(y^1,t)\Big\}\psi(x), 
~~~\gg^\mu\delmu\psi(x) = 0.
\label{TWsol} 
\eeq
Product of two fermion operators has to be regularized by a   
point-splitting. The integral in the exponent contributes naturally to 
find ($j^\mu$ and $j^\mu_5$ are the free currents)  
\beq   
J^\mu(x) = j^\mu(x) - \frac{e}{\gp}B^\mu(x),~~ 
J_5^\mu(x) = j_5^\mu(x) - \frac{e}{\gp}\ge^{\mu\nu}B_\nu(x).  
\label{Bcur}
\eeq   
Inserting the above $J^\mu(x)$ to the Proca equation, one finds that the 
the bare mass is replaced by $\mu^2 = \mu_0^2+e^2/\gp$ and that this equation  
can be easily inverted since only the free fields are involved. 
Thus, there is no dynamically independent vector field. Following 
our method, we insert the solutions for $B^\mu$ and $\Psi$ into 
Lagrangian and then derive the Hamiltonian. The question if the latter 
will be diagonal or will have to be diagonalized, together with other 
properties of the model, is under study.   

\section{The Schwinger model in the Landau gauge}  
\label{sec:4} 
The masslessness of the vector field makes the Schwinger model more subtle 
than was the previous one. The key question is to correctly handle the 
gauge variables since in the covariant 
gauge $\delmu A^\mu = 0$ not all gauge freedom has been removed. We 
implement the gauge condition in the Lagrangian as \cite{kh}:
\beq   
{\cal L}=\frac{i}{2}\Psib\gg^\mu\delrlmu \Psi-\frac{1}{4}F_{\mu\nu}F^{\mu\nu} 
-eJ_\mu(x) A^\mu(x) -G(x)\delmu A^\mu(x) + \hlf(1-\gg)G^2(x),~~~ 
F_{\mu\nu}=\delmu A_\nu -\delnu A_\mu.
\label{SML}
\eeq   
The gauge-fixing terms furnish  
the component $A^0$ with the conjugate momentum, $\gP_{A^0}(x) = - G(x)$.  
Moreover, they guarantee restriction to an arbitrary 
covariant gauge in which neither the condition $\delmu A^\mu(x)=0$ nor the 
Maxwell equations $\delmu F^{\mu\nu}(x) = eJ^\nu(x)$ 
hold as operator relations. The gauge fixing field obeys 
$\delmu\partial^\mu G(x) =0$ so that positive and negative-frequency parts  
$G^{(\pm)}(x)$ are well defined. 
%+++Note also that the gauge-fixing term furnishes  
%+++the component $A^0$ with the conjugate momentum, $\gP_G(x) = - G(x)$.  

Our strategy is to proceed in the spirit 
of K. Haller's generalization \cite{kh} of the Gupta-Bleuler quantization, in  
which the unphysical components of the gauge field are represented as ghost 
degrees of freedom of zero norm, carrying vanishing momentum and energy. To 
ensure that we are dealing with the original 2-dimensional QED, we have 
to restrict the theory to the physical subspace $G^{(+)}
\vert phys \rangle=0$.   

We will choose $\gg=1$ in the above Lagrangian. Then the gauge condition is an operator  
relation while the (modified) Maxwell and Dirac equations read 
%+++ and the solution of the Dirac equation  
\beq
\delmu F^{\mu\nu}(x) = eJ^\nu(x) - \partial^\nu G(x),~~~
i\gg^\mu\partial_\mu\Psi(x)= e\gg_\mu A^\mu(x)\Psi(x).  
\label{Adir}
\eeq
The solution of the latter is completely analogous to the Thirring-Wess model 
case, Eq.(\ref{TWsol}). Again, the vector 
and axial-vector currents have to be calculated via point-splitting with an 
important difference: the exponential of the line integral of the gauge field 
must be inserted in the current definition to compensate for the violation 
of gauge invariance due to the point splitting. After inserting the 
calculated interacting current into the Maxwell equations,    
we have to express the Lagrangian and Hamiltonian in terms of the free fields 
as before. The physical picture will become transparent if we   
make a unitary transformation to the Coulomb-gauge representation \cite{kh}. 
Before performing this step, let us indicate the main   
ingredients of the original covariant-gauge solution \cite{LSw,AAR} and point 
out a problem with it. 

The starting point was the Ansatz for the gauge field and the currents, 
($\tilde{\delmu}=\ge_{\mu\nu}\partial^\nu$) 
\beq
A_\mu = -\frac{\sqrt{\gp}}{e}\big(\tilde{\partial}_\mu\Sigma +
\delmu\tilde{\eta}\big),~~~ 
J^\mu = \Psib\gg^\mu\Psi = -\frac{1}{\sqrt{\gp}}\tilde{\partial}^\mu 
\Phi,~~~
J_5^\mu = \Psib\gg^\mu\gg^5\Psi=-\frac{1}{\sqrt{\gp}}\partial^\mu\Phi,  
\label{ansa}
\eeq 
where $\gS, \tilde{\eta}$ and $\Phi$ are so far unspecified scalar fields. 
In the $\delmu A^\mu = 0$ gauge, one finds $\delmu\partial^\mu \tilde{\eta}=0$ 
and 
$F_{\mu\nu} = \frac{\sqrt{\gp}}{e}\ge_{\mu\nu}\partial_\gr\partial^\gr \gS.$ 
%+++Assu $J^\mu = \Psib\gg^\mu\Psi = -\frac{1}{\sqrt{\gp}}\tilde{\partial}^\mu 
%+++\Phi$,~
%+++$J^\mu = \Psib\gg^\mu\gg^5\Psi=-\frac{1}{\sqrt{\gp}}\partial^\mu\Phi$,   
From the anomalous divergence of the axial current
\beq
\delmu J^\mu_5 = \frac{e}{2\gp}e_{\mu\nu}F^{\mu\nu} 
\label{axano}
\eeq
one concludes that $\delmu\partial^\mu \Phi = \delmu\partial^\mu \gS$ or 
$\Phi = \gS +h$ with the free massless field $h$ obeying  
$\delmu\partial^\mu h = 0$. Then the vector current is
\beq
J_\mu = -\frac{1}{\sqrt{\gp}}\tilde{\partial}_\mu + L_\mu,~~
L_\mu= -\frac{1}{\sqrt{\gp}}\tilde{\partial}_\mu h.
\label{nc}
\eeq
From the Maxwell eqs.
\beq
\tilde{\partial}^\nu\big(\partial_\gr\partial^\gr + \frac{e^2}{\gp}\gS\big) - 
\frac{e^2}{\sqrt{\gp}}L^\nu = 0
\label{meq}
\eeq
one concludes that $\tilde{\partial}^\mu L_\mu = 0$ or
\beq
\big(\partial_\gr\partial^\gr + \frac{e^2}{\gp}\big)\gS = 0.
\label{mass}
\eeq
One component of the gauge field, namely $\gS(x)$, became massive. For 
consistency reasons, $L_\mu$ can vanish only weakly,   
$\langle\psi\vert L_\mu(x)\vert \psi\rangle =0$.  
With the above Ansatz for $A^\mu$, Dirac eq. becomes 
%+++($\ge^{\mu\nu}
%+++\gg_\nu = \gg^\mu\gg^5$)  
$i\gg^\mu\delmu\Psi = -\sqrt{\gp}\gg^\mu\gg^5\delmu\big(\gS+\eta\big)\Psi$ 
with the solution
\beq
\Psi(x) = :e^{i\sqrt{\gp}\gg^5\big(\gS(x)+\eta(x)\big)}:\psi(x).
\label{ssol}
\eeq
Calculation of the currents via the point-splitting yields identification 
$h(x)=\eta(x) +\varphi(x)$, where $\varphi(x)$ is the "potential" (integrated 
current) of the free currents.
From $[A^1(x^1),\gp_1(y^1)]=i\gd(x^1-y^1)$ and with  
$\gp_1=F^{01} = 
\frac{e}{\sqrt{\gp}}\gS$ one gets the equal-time commutator      
$[\gS(x^1),\partial_0\gS(y^1)]=i\gd(x^1-y^1)$ of a canonical scalar field.
Inserting Eq.(\ref{ssol}) to the original Lagrangian, we  
derive the physical part of the Hamiltonian as : 
\beq
H = \intex \Big[-i\psi^\dagger\ga^1\partial_1 \psi + \frac{1}{2}\frac{e^2}
{\gp}\gS^2\Big].
\label{bh}
\eeq
This Hamiltonian is non-diagonal when expressed in terms of Fock  
operators of $\gS(x)$. A Bogoliubov transformation is  necessary.  
A coherent-type of vacuum state will be obtained as the true vacuum.  

An interesting aspect of the model is its vacuum degeneracy and the theta 
vacuum. In the work \cite{LSw} the mechanism generating multiple 
vacua is based on "spurion" operators. These however have been 
shown to be an artifact of the incorrect treatment of residual gauge freedom 
\cite{MStr}. What can be the true mechanism of the vacuum degeneracy in the 
Scwinger model? We believe that it is the presence of a gauge zero mode in 
the finite-volume treatment of the model \cite{HH} together with a quantum 
implementation of residual invariance under {\it large} gauge transformations 
as desribed in \cite{lmlf}. This leads us naturally to the finite-volume 
reformulation of our approach outlined in the first part of this section.   

\section{Spontaneous symmetry breaking in LF theory with fermions}
\label{sec:5} 
At the LC workshop in Valencia, Marvin Weinstein criticised the way how the LF 
theory describes spontaneous symmetry breaking, saying that Goldstone (or 
hidden) symmetry is a chalenge for LF theory. Where is vacuum degeneracy? 
Actually the latter can be described if one takes into account a mechanism 
based on the presence of dynamical fermion zero modes \cite{lmssb}. 
$O(2)$-symmteric sigma model provides us with a good example. 
Its Hamiltonian is symmetric under axial-vector transformations 
\beq
\Psi_+(x) \rightarrow e^{-i\beta\gg^5}\Psi_+(x) = V(\gb)\Psi_+(x)V^{-1}(\gb)
,~~V(\gb) = \exp\{-i\gb Q_5\},~~Q_5 = \int_V d^3\ulix J^+_5(x).  
\label{xt}
\eeq
The operator of the axial charge will not annihilate the LF vacuum since 
in addition to the normal-mode part (which annihilates it) it contains also 
the zero-mode term,  $Q^5 = Q^5_N + Q^5_0$, where  
\bea 
Q^5_0 = \sum_{p_\perp,s}2s\Big[\Big(b_0^\dagger(p_\perp,s)
d_0^\dagger(-p_\perp,-s) + H.c. \Big) +  
%+++ + b_0(p_\perp,s)d_0(-p_\perp,-s) + 
b_0^\dagger(p_\perp,s)b_0(p_\perp,s) -  
d^\dagger_0(p_\perp,s)d_0(p_\perp,s)\big].
\label{FQ5}
\eea 
%+++when the $J^+(x)$ current integrated over the volume, the $p^+=0$ component 
%+++of the charge maintains all four operator structures including 
The term $b_0^\dagger(p_\perp)d_0^\dagger(-p_\perp)$ will   
generate an infinite set of degenerate vacuum states.  
One has all properties for deriving the Goldstone theorem in the usual way. 

%relationship to the continuum results of F.Coester and W.Polyzou?

{\bf We conclude} {\sf  with the statement that the realm of exactly solvable models 
%+++contains certain inconsistencies and aspects that are not completely under 
still offers us certain surprises and room for improvement. And that degenerate vacua exist in the LF formalism in spite of its kinematically defined vacuum 
state.} 
%+++control while at the same time offers us a new, sometimes rather surprising 
%+++physical picture. 

{\bf Acknowledgements} This work has been supported by the grant VEGA No. 
2/0070/2009 and by the Slovak CERN Commission. The author also thanks 
Pierre Grang\'e for fruitfull discussions.

\end{document}